\newcommand{\beq}{\begin{equation}}
\newcommand{\eeq}{\end{equation}}
\newcommand{\sh}{{\rm sh}}
\newcommand{\ch}{{\rm ch}}
\newcommand{\id}
 {i\kern.06em\hbox{\raise.25ex\hbox{$/$}\kern-.60em$\partial$}}
\newcommand{\as}{\!\not\!\! A}

\newcommand{\rD}{\!\not\!\! D}
\newcommand{\rbD}{\!\not\!\! {\bf D}}
\newcommand{\bs}{/\kern-.52em b}
\newcommand{\ds}{/\kern-.52em d}
\newcommand{\qs}{/\kern-.52em s}
\newcommand{\D}{{\cal D}}

\newcommand{\cL}{{\cal L}}

\newcommand{\p}{\partial}
\newcommand{\yp}{^{\prime}}
\newcommand{\bpsi}{\bar{\psi}}

\newcommand{\dd}
{\kern.06em\hbox{\raise.25ex\hbox{$/$}\kern-.60em$\partial$}}

\newcommand{\ep}{\epsilon}

\documentstyle[12pt]{article}
\date{}
\begin{document}
\title{ (2+1)-Dimensional Fermion Determinant in a Constant
Background Field at Finite Temperature and Density
\footnotetext{\# Corresponding author}
\thanks{S.S. Feng is on leave of absence from the Physics Department of Shanghai
University, 201800, Shanghai, China}}
\author{{Sze-Shiang Feng$^{1,2,\#}$, Zi Yao$^3$, De-Pin Zhao$^3$,
 De-Si Zang$^3$}\\
1.{\small {\it CCAST(World Lab.), P.O. Box 8730, Beijing 100080}}\\
2.{\small {\it Department of Astronomy and Applied Physics}}\\
  {\small {\it University of Science and Technology of China
            230026, Hefei, China}}\\e-mail:sshfeng@yahoo.com\\
3. {\small {\it Physics Division, Anhui Post \& Telecommunication
    School,230031, Hefei, China}}}
\maketitle
\newfont{\Bbb}{msbm10 scaled\magstephalf}
\newfont{\frak}{eufm10 scaled\magstephalf}
\newfont{\sfr}{eufm7 scaled\magstephalf}
\baselineskip 0.23in
\begin{center}
\begin{minipage}{135mm}
\vskip 0.3in
\baselineskip 0.23in
\begin{center}{\bf Abstract}\end{center}
  {We evaluate exactly both the non-relativistic and relativistic
  fermion determinant in 2+1 dimensions in a constant background
  field at finite temperature. The effect of finite chemical potential
  is also considered. In both cases, the systems are decoupled
  into an infinite number of 1+1 fermions by Fourier transformation
  in the $\beta$-variable. The total effective actions
  demonstrate non-extensiveness in the $\beta$ dimension.
  \\PACS number(s):11.10.Wx,11.30.Er
   \\Key words: fermion determinant, chemical potential, temperature}
\end{minipage}
\end{center}
\vskip 1in
\section{Introduction}
 Thanks to the exotic mathematical structure and the possible
 relevence to condensed matter
physics in two space dimensions, Chern-Simons(CS) models
have drawn much attention in the past decade
\cite{s1}\cite{s2}(For a review, see \cite{s3}).The CS term  can be either put in by
hand , or more naturaly, induced
by fermion degrees, as a part of the original (effective) lagrangian.Two properties
of the CS action are fundamental. One is that it is odd under parity transform due
to the presence of three dimensional Levi-Civita symbol. The other is that it is invariant
under {\it small} gauge transforms while non-invariant under {\it large }
gauge transforms (those not to be continuously deformed to unity and thus carrying non-trivial winding
numbers)\cite{s4}. 
In the free spacetime whose topology is trivial, the homotopy group $\pi_3$
is trivial 
in the Abelian case. But there may be nontrivial 
 large gauge transformations  if the gauge
fields are
subject to non-trivial boundary
 conditions(for a more recent discussion  see\cite{s5}).
In general, if there exists non-trivial $\pi_3$, the quantum theory is consistent only
if the CS parameters are quantized. There then arises a problem:what happens to
the quantized parameters by quantum corrections? In the zero temperature, the induced
CS term is well-understood \cite{s6}-\cite{s10}. 
But at finite temperature,
it was argued \cite{s11}that the coefficient of the CS term in the
effective action for the gauge field should remain unchanged at finite temperature.
Yet,  a naive perturbative
calculation that mimics that at zero temperature leads to a CS term with a parameter
continuously dependent on the temperature\cite{s11}-\cite{s12}. Therefore, the behavior
under gauge transforms seems to be temperature-dependent. 
The problem of quantum corrections to the CS coefficient induced by fermions at finite
temperature was re-examined in\cite{s13}
, where it was concluded that, on gauge invariance grounds and in perturbation
theory, the effective action for the gauge field can not contain a smoothly renormalized CS coefficient at
non-zero temperature. Obviously, it is neccessary to obtain some exact result
in order to reconcile the contradiction.
As a toy model, the effective action
of a (0+1) analog of the 2+1 CS system was exactly calculated
\cite{s14}. It shows that in the analog, the exact finite $T$ effective action
, which is non-extensive in temperature,
has a well-defined behavior under a large gauge transformation,{\it
independent
 of the temperature}, even though at any given finite order of 
a perturbation expansion, there is a temperature dependence.
So it implies that the discussions of the gauge invariance
of finite temperature effective actions and induced CS terms
in higher dimensions requires consideration of the full
perturbation series. Conversely, no sensible conclusions
may be drawn by considering only the first finite number
of terms in the expansion. The course of being exactly calculable
is that the gauge field can be made constant
by gauge-transformations. Employing this trick,
Fosco {\it et.al.} calculated exactly the parity breaking part of the
fermion determinant
in 2+1 dimensions with a particular background gauge field.
for both Abelian and non-Abelian cases\cite{s15}\cite{s16},
and the result agrees with that from the $\zeta$-function methed
\cite{s5}. 
More general background gauge fields were also considered\cite{s17}.
All these works show that (restricted to that particular {\it ad hoc}
configuration) gauge invariance of the
effective action
is respected even when large gauge transformations are considered.
 It is now clear that the effective action induced
by the fermion determinant is in general a non-extensive quantity
in space-time/temperature and
this feature enables the effective action preserve gauge invariance.\\
\indent In the non-relativistic case, The effective action induced by
the 2+1 fermion determinant was studied in\cite{s18}
perturbatively in order to investigate the possible relevence between
Chern-Simons
theory and superconductivity, at both zero and finite temperature.
Since  the determinant can not be evaluated exactly for general
background gauge fields,
ref.\cite{s19} considered the case that the gauge field is that of a
constant magnetic field and discussed the induced quantum numbers.
The difference between the perturbative (loop) calculations and the rigorous
results in this special case is demonstrative. \\
\indent The effect of  finite chemical potential should be taken into account
whenever discussing the statistical physics of a grand canonical ensemble.
It was shown that in 1+1 dimensions, the non-zero chemical potential may contribute
a non-trivial phase factor to the partition function\cite{s20} .
The problem for an arbitrary background in 2+1 dimensions was tackled 
perturbatively in \cite{s21}. As ususal, gauge transform property
of the effective action suffers some temperature-dependence. Using the
same technique as in \cite{s15}, the effect on the parity-odd
part of non-zero chemical
potential is considered in\cite{s22} but the parity-even part
can not be obtained exactly for the background therein. Therefore,
it is worthwhile considering the problem by exact computation with
some particular background. This is the topic of this paper.
 The layout of this paper is as follows.
In sections 2 and 3 we exactly evaluate the non-relativistic and relativistic
fermion
determinant at finite temperature and finite density in a constant magnetic
field. Section 4 is devoted to conclusional discussions.
\section{The non-relativistic case}
 The fermion Lagrangian is\cite{s23}
\beq
\cL=\psi^{\dag}iD_0\psi-\frac{1}{2m}\psi^{\dag}\rbD^2\psi
\eeq
where $\rbD=\gamma^iD_i, i=1,2. D_{\mu}=\p_{\mu}+ieA_{\mu},
\gamma^0=\sigma_3, \gamma^1=i\sigma_2, \gamma^2=i\sigma_1.
e=-|e|$,  and $\sigma_{1,2,3} $ are the usual three Pauli matrices.
We choose representation of gamma matrices so that it gives the correct
sign of the Zeeman energy. It can be calculated directly that
\beq
\frac{1}{2m}\rbD^2=\frac{1}{2m}(D_iD^i+\frac{1}{4}[\gamma^i, \gamma^j]
ieF_{ij})
=\frac{1}{2m}(-{\bf D}^2)-g_s\mu_BBs_z
\eeq
where $\mu_B=e/2m, g_s=2$ is the electron $g$-factor for spin.
 We incorporate an external filed $b$ in order to discuss the
spin. 
The Euclidean action at finite temperature and finite density
reads then
\beq
\cL_E=\psi^{\dag}[D_{\tau}-\frac{1}{2m}\rbD^2
-bs_z+\mu]\psi
\eeq
The effective action $\Gamma$ is given by definition
\beq
e^{-\Gamma}=\int_{A.B.C.}\D\psi^{\dag}\D\psi
\exp\{-\int^{\beta}_0 d\tau\int d^2{\bf x}[
\psi^{\dag}D_{\tau}\psi-\frac{1}{2m}\psi^{\dag}\rbD^2
\psi+\mu\psi^{\dag}\psi-b\psi^{\dag}s_z\psi]\}
\eeq
where the $A.B.C.$ implies that the functional integral over the
fermion fields is implemented with anti-periodic boundary conditions.
Once $\Gamma$ is known, the induced particle number and
spin are provided by
\beq
<N>=\int d^2{\bf x}<\psi^{\dag}\psi>=-\frac{1}{\beta}\frac{\p}{\p\mu}\Gamma
_{b=0}
\eeq
\beq
\int d^2{\bf x}<\psi^{\dag}s_z\psi>=-\frac{1}{\beta}\frac{\p}{\p b}\Gamma
_{b=0}
\eeq
where we have used the fact that the correlation functions 
in (5) and (6) are actually $\beta$-independent.\\
\indent Since the exact evaluation of the functional integration in (4)
in general is beyond our ability so far, we first consider those
backgrounds of the following spacetime dependence as in\cite{s15}
\beq
A_{\tau}=A_{\tau}(\tau), ~~~~~~~~~ A_i=A_i({\bf x})
\eeq
Then the gauge field can be rendered constant in "time" $\tau$ by
gauge transformations. The time component $A_{\tau}$ will be its
average value $\tilde{A}_{\tau}=\beta^{-1}\int^{\beta}_0 d\tau
A_{\tau}(\tau)$, which can not be transformed away by {\it local}
transformations. In this gauge, we can employ the Fourier transformation
\beq
\psi(\tau,{\bf x})=\beta^{-1/2}\sum^{+\infty}_{-\infty}\psi_n({\bf x})
e^{i\omega_n\tau}, ~~~~~~~~~~
\psi^{\dag}(\tau,{\bf x})=\beta^{-1/2}\sum^{+\infty}_{-\infty}\psi_n^{\dag}
({\bf x})e^{-i\omega_n\tau}
\eeq
where $\omega_n=\frac{(2n+1)\pi}{\beta}$, to decouple the system as a
sum of an infinite number of fermions in 1+1 dimensions.
\beq
e^{-\Gamma}=\prod_n\int\D\psi^{\dag}_n({\bf x})\D\psi_n({\bf x})
\exp\{-\int d^2{\bf x}\psi^{\dag}_n({\bf x})[(i\omega_n+ie\tilde{A}
_{\tau})
-\frac{1}{2m}(-{\bf D}^2)-b^{\yp}s_z+\mu]\psi_n({\bf x})\}
\eeq
($b^{\yp}=b-g_s\mu_B B$) where we have
used the transformation of the functional measure
\beq
\D\psi^{\dag}(\tau, {\bf x})\D\psi(\tau, {\bf x})=\prod_n
\D\psi^{\dag}_n({\bf x})\D\psi_n({\bf x})
\eeq
which can be easily proved from the orthonormality of the basis
$\{\beta^{-1/2}e^{i\omega_n\tau}\}$ in the Fourier transformation.
It can be seen easily that once the eigenvalues of the operator
$\frac{1}{2m}(-D^2)+b^{\yp}s_z$
are known, the functional integration can be accomplished
readily. Unfortunately, this is impossible for general
gauge field backgrounds, even for the restricted class
(7). Therefore, we need to make further restrictions. The simplest
case is that the magnetic field $F_{ij}$ is constant $F_{12}=B$ and the
the corresponding gauge potential can be chosen in the gauge
${\bf A}=(-By,0)$. In this case, the eigenvalues of the operator
$\frac{1}{2m}(-D^2)+b^{\yp}s_z$
can be acquired from the solutions of the equation
\beq
\{\frac{1}{2m}[(P_x+eBy)^2+P_y^2]+b^{\yp}s_z\}
\chi=\lambda\chi
\eeq
where $\chi$ is a two-component spinor and $P_i=-i\p_i$
. The solutions to (11) are easy
to find and the eigenvalues can be obtained from the well-known Landau
levels,i.e.
\beq
\lambda_{l,s_z}=(l+\frac{1}{2})\Omega+b^{\yp}s_z
~~~~~~~~l=0,1,2,\cdots; s_z=\pm\frac{1}{2};~~~~~~~\Omega=\frac{|eB|}{m}
\eeq
These energy levels are highly degenerate with degeneracy $\frac{|eB|}{2\pi}$
per unit area
which must be taken into account when calculating the fermion
determinant.
\beq
e^{-\Gamma}=\prod_n{\rm Det}[i\omega_n+ie\tilde{A}_{\tau}
 -\frac{1}{2m}(-{\bf D}^2)-b^{\yp}+\mu]
\eeq
There is one important point that deserves attention here. In the absence
of external magnetic field, the Hamiltonian is just that of a free
electron and the energy eigenvalue spectrum is continuous which can not be
regarded simply as the limit of the discrete spectrum 
for vanishing external field. Since
the numerator is
\beq
{\rm Det}[i\omega_n+ie\tilde{A}_{\tau}
 -\frac{1}{2m}(-{\bf D}^2)-b^{\yp}s_z+\mu]
=\{\prod_{l=0}^{\infty}\prod_{s_z=\pm\frac{1}{2}}[i\omega_n
-E_{l\pm}+\mu]\}^{\frac{|eB|}{2\pi}}
\eeq
\beq
E_{l\pm}=ie\tilde{A}_{\tau}
+(l+\frac{1}{2})\Omega\pm\frac{b^{\yp}}{2}
\eeq
we have
\beq
-\Gamma=\frac{|eB|}{2\pi}\sum_l\sum_n[\ln(i\omega_n-E_{l+}
+\mu)+\ln(i\omega_n-E_{l-}+\mu)]
\eeq
Using the formula for fermion\cite{s24}
\beq
\sum_n\frac{1}{i\omega_n-x}=\frac{\beta}{e^{\beta x}+1}
\eeq
We have then the expectation values of the spin-up and spin-down electrons
per unit area
\beq
N_{\pm}=\frac{|eB|}{2\pi}\sum_l\frac{1}{e^{\beta(E_{l\pm}-\mu)}+1}
\eeq
\beq
<s_z>=\frac{1}{2}(N_+-N_-)
\eeq
\beq
M_z=g_s\mu_B<s_z>
\eeq
At zero temperature, these results coincide with those of
\cite{s19}
\section{The relativistic case}
\indent The Lagrangian of the fermion is
\beq
\cL=\bpsi(i\gamma^{\mu}\rD_{\mu}-m)\psi
\eeq
There are two inequivalent representations of the $\gamma$-matrices
in three dimensions:$\gamma^{\mu}=(\sigma_3, i\sigma_2, i\sigma_1)$
and $\gamma^{\mu}=(-\sigma_3, -i\sigma_2, -i\sigma_3)$. We choose the first.
As usual, the total effective action $\Gamma(A,m,\mu)$ at finite temperature
is defined as
\beq
e^{-\Gamma(A,m,\mu)}=\int\D\psi\D\bpsi\exp[-\int^{\beta}_0 d\tau\int d^2x\bpsi
(\dd+ie\as+m-\mu\gamma^3)\psi]
\eeq
where we are using Euclidean Dirac matrices in the representation
 $\gamma_{\mu}=(\sigma_3, \sigma_2,\sigma_1)
 $, and $\beta$ is the inverse temperature. It makes no difference
whether the indices are lower or upper. The label 3 refers actually to the
Euclidean time component. The fermion fields are subject to antiperiodic
boundary conditions while the gauge field are periodic. Under parity
transformation, 
\beq
x^1\rightarrow-x^1,x^2\rightarrow x^2, x^3\rightarrow x^3;\psi\rightarrow\gamma^1\psi,
\bpsi\rightarrow -\bpsi\gamma^1; A^1\rightarrow -A^1,A^2\rightarrow A^2,A^3\rightarrow A^3
\eeq
($\gamma$ matrices are kept intact). So only the mass term varies under the
parity transformation. As in \cite{s15}, the parity-odd part  is defined
as
\beq
2\Gamma(A,m,\mu)_{\rm odd}=\Gamma(A,m,\mu)-\Gamma(A,-m,\mu)
\eeq
It is not an easy task to calculate (22) for general configuration of the gauge field.
A particular class of configurations of $A$ for which (22) can be exactly
computed is that defined by (7).
This class of gauge fields shares the same feature as in the 0+1 dimensions
in\cite{s14}: the time
dependence of the time component can be erased by gauge transformations. Therefore, the Euclidean
action can be decoupled as a sum of an infinite 1+1 actions
\beq
e^{-\Gamma(A,m,\mu)}=\int\D\psi_n(x)\D\bpsi_n(x)\exp\{-\frac{1}{\beta}\sum^{+\infty}_{-\infty}
\int d^2x\bpsi_n(x)[\ds+m+i\gamma^3(\omega_n+e\tilde{A}_3)-\mu\gamma^3]\psi_n(x)\}
\eeq
where $\ds=\gamma_j(\p_j+ieA_j)$ is the 1+1 Dirac operator and 
$\tilde{A}_3$ is the
mean value of $A_3(\tau)$. It is seen that the chemical potential in 2+1 dimensions plays the
role of a chiral potential in 1+1 dimensions. Let us introduce $\Omega_n$ for convenience,
$\Omega_n=\omega_n+e\tilde{A}_3$. Since
\beq
m+i\gamma^3\Omega_n-\mu\gamma^3=\rho_ne^{i\gamma_3\phi_n}
\eeq
where
\beq
e^{2i\phi_n}=\frac{m-\mu+i\Omega_n}{m+\mu-i\Omega_n}
\eeq
and
\beq
\rho_n=\sqrt{(m+\mu-i\Omega_n)(m-\mu+i\Omega_n)}
\eeq
we have therefore
\beq
{\rm det}(\dd+ie\as+m-\mu\gamma^3)=\prod^{+\infty}_{n=-\infty}
{\rm det}[\ds+\rho_ne^{i\gamma_3\phi_n}]
\eeq
Explicitly, the 1+1 determinant for a given mode is a functional integral
over 1+1 fermions
\beq
{\rm det}[\ds+\rho_ne^{i\gamma_3\phi_n}]=\int\D\chi_n\D\bar{\chi}_n
\exp\{-\int d^2x\bar{\chi}_n(x)(\ds+\rho_ne^{i\gamma_3\phi_n})\chi_n(x)\}
\eeq
After implementing a chiral rotation whose Jaccobian is wellknown (the Fujikawa
method applies also to complex chiral parameters), we obtain
\beq
{\rm det}[\ds+m+i\gamma^3(\omega_n+e\tilde{A}_3)-\mu\gamma^3]=J_n{\rm det}
[\ds+\rho_n]
\eeq
where
\beq
J_n=\exp(-i\frac{e\phi_n}{2\pi}\int d^2x \ep^{jk}\p_jA_k)
\eeq
Note that the chiral anomalies, or the Jaccobian$J$, dependes on the
boundary conditions as well. If the system is defined on a torus and the
fields are subject to periodic boundary conditions, for instance
,$A_j(x,y)=A_j(x+L_x,y),A_j(x,y)=A_j(x, y+L_y)$, the trace of $\gamma_5$ in
\cite{s25}
is taken over discrete complete set instead of the continuous
plane waves. Thus the momentum integral $\int
\frac{d^2k}{(2\pi)^2}e^{-k^2}=\frac{1}{4\pi}$ should be replaced by
$\frac{1}{L_xL_y}\sum_{n_1,n_2}\exp[-(\frac{2\pi}{L_x}n_1)^2-(\frac
{2\pi}{L_2}n_2)^2]$.
 Using the formula $\sum^{+\infty}_{n=_\infty}
e^{-\pi
zn^2}=\frac{1}{\sqrt{z}}\sum^{+\infty}_{n=-\infty}e^{-\frac{\pi}{z}
n^2}$\cite{s26}
which holds for any complex $z$ with $Rez>0$. 
we have
\beq
\frac{1}{L_xL_y}\sum_{n_1,n_2}\exp[-(\frac{2\pi}{L_x}n_1)^2-(\frac
{2\pi}{L_2}n_2)^2]=\theta(L_x)\theta(L_y)
\eeq
where
$\theta(L)=\frac{1}{\sqrt{4\pi}}\sum^{+\infty}_{n=-\infty}e^{-\frac{L^2}
{4}n^2}
$. In this case, (32) should be replaced by
\beq
J_n=\exp(-2ie\phi_n \theta(L_x)\theta (L_y)\int_{L_x\times L_y} d^2x
\ep^{jk}\p_jA_k)
\eeq
In the following, we only concentrate on the infinite space case since
 the conclusion on a torus can be obtained by a trivial substitution.
Fortunately, we also have  $\rho_n(m)=\rho_n(-m)$ for finite chemical potential. Thus 
 we have
immediately
\beq
\Gamma_{\rm odd}=-\sum^{+\infty}_{n=-\infty}\ln J_n=
i\frac{e}{2\pi}\sum^{+\infty}_{n=-\infty}\phi_n\int d^2x \ep^{jk}\p_jA_k
\eeq
To calculate $\sum^{+\infty}_{n=-\infty}\phi_n$, we need to compute
$\prod^{+\infty}_{n=-\infty}\frac{m-\mu+i\Omega_n}{m+\mu-i\Omega_n}
$.
  Using the formula $\prod_{n=1,3,5,..}[1-\frac{4a^2}{(2n-1)^2}]=\cos\pi a$
as in \cite{s1}, we have ($a=e\tilde{A}_3$)
\beq
\prod^{+\infty}_{n=-\infty}e^{2i\phi_n}=
\prod^{+\infty}_{n=-\infty}\frac{m-\mu+i\Omega_n}{m+\mu-i\Omega_n}
=\frac{\ch\frac{\beta}{2}(m-\mu)+i\sh\frac{\beta}{2}(m-\mu){\rm tg}\frac{\beta a}{2}}
      {\ch\frac{\beta}{2}(m+\mu)-i\sh\frac{\beta}{2}(m+\mu){\rm tg}\frac{\beta a}{2}}
\eeq
Therefore
\beq
\Gamma_{\rm odd}=\frac{e}{4\pi}\ln
[\frac{\ch\frac{\beta}{2}(m-\mu)+i\sh\frac{\beta}{2}(m-\mu){\rm tg}\frac{\beta a}{2}}
      {\ch\frac{\beta}{2}(m+\mu)-i\sh\frac{\beta}{2}(m+\mu){\rm tg}\frac{\beta a}{2}}
]\int d^2x\ep^{jk}\p_jA_k
\eeq
which is quite different from the perturbative conclusion in \cite{s19}. 
(The formula eq(97) there is for an arbitrary background). \\
\indent Now the low temperature limit can be
obtained. It will
depend on
the
relationship between $m$ and $\mu$.\\ 
(i).If $m>\mu,m+\mu>0$
\beq
\lim_{\beta\rightarrow\infty}\Gamma_{\rm odd}=\frac{e}{4\pi}
\beta(ia-\mu)\int d^2x \ep^{jk}\p_jA_k
\eeq
(ii). If $m-\mu>0, m+\mu<0$,
\beq
\lim_{\beta\rightarrow\infty}\Gamma_{\rm odd}=\frac{e}{4\pi}
\beta m\int d^2x \ep^{jk}\p_jA_k
\eeq
(iii).If $m<\mu,m+\mu>0$
\beq
\lim_{\beta\rightarrow\infty}\Gamma_{\rm odd}=\frac{e}{4\pi}
(-\beta m)\int d^2x \ep^{jk}\p_jA_k
\eeq
(iv).If $m<\mu,m+\mu<0$
\beq
\lim_{\beta\rightarrow\infty}\Gamma_{\rm odd}=\frac{e}{4\pi}
\beta(\mu-ia)\int d^2x \ep^{jk}\p_jA_k
\eeq
(v).If $m=\mu$
\beq
\lim_{\beta\rightarrow\infty}\Gamma_{\rm odd}=\frac{e}{4\pi}
(-\beta m+i\frac{\beta a}{2})\ln\cos\frac{\beta a}{2}\int d^2x
\ep^{jk}\p_jA_k
\eeq
 It vanishes in the high temperature limit. It is obvious that the low temperature
is very sensitive to the values of $m$ and $\mu$, as agrees with the results perturbatively
obtained \cite{s19}.\\
\indent Since in the large-$m$ limit (or in the low-density limit), the
parity-odd part dominantes over the effective
action, and the particle number in the ensemble is
$<N>=\frac{1}{\beta}\frac{\p}{\p\mu}\ln Z(\beta,\mu)$, we have from the limits 
(38) and (41)
that the flux should be quantized\cite{s27},
\beq
\Phi=<N> \frac{4\pi\hbar}{e}
\eeq
which implies that each particle carries flux $\frac{4\pi\hbar}{e}$
and thus should be of fractional spin $S_{\otimes}=\frac{1}{4}$.
This is
in accordance with the conclusion in \cite{s19}.\\
\indent The next thing is to evaluate ${\rm det}(\ds+\rho_n)$. We have to
calculate the eigenvalues of the operator $\ds+\rho_n$ for this purpose,
i.e. to solve the equation
\beq
(\ds+\rho_n)\psi=\lambda\psi
\eeq
In general, it is impossible to solve it. So we
confine ourselvs to the background (7).
It is easily seen that once the eigenvalues of $\ds$ are known,
the eigenvalues $\lambda$ can be obtained. We thus consider the problem
\beq
\ds\psi=a\psi
\eeq
Since
\beq
\ds^2=D_iD_i\frac{1}{4}[\gamma_j,\gamma_i][D_j,D_i]=D_iD_i+eB\sigma_3
\eeq
the eigenvalues $a$ can be obtained from the well-known relativistic
Landau levels\cite{s28}.
\beq
a=\pm i\sqrt{2(l+\frac{1}{2})|eB|-2eBs_{\pm}}
\eeq
with degeneracy $\frac{|eB|}{2\pi}$ per unit area. Accordingly, we have
\beq
\lambda_{l,s_{\pm}}=\rho_n\pm i\sqrt{2(l+\frac{1}{2})|eB|-2eBs_{\pm}}
\eeq
Therefore, we have (we suppose $eB>0$) for a unit area
\begin{eqnarray}
{\rm det}(\ds+\rho_n)&=&\frac{|eB|}{2\pi}
\prod_{l=0}^{\infty}[\rho_n+i\sqrt{2(l+\frac{1}{2})|eB|-eB}]
[\rho_n-i\sqrt{2(l+\frac{1}{2})|eB|+eB}]\\
&=&\frac{|eB|}{2\pi}
\rho_n\prod_{l=o}^{\infty}[\rho_n+i\sqrt{2(l+1)eB}][\rho_n
-i\sqrt{2(l+1)eB}]\\
&=&\frac{|eB|}{2\pi}\rho_n\prod_{l=0}^{\infty}(\rho^2+2(l+1)|eB|)
\end{eqnarray}
Another way to evaluate it is to make use of the relation
\beq
{\rm det}(\ds+\rho_n)={\rm det}[\sigma_3(\ds+\rho_n)\sigma_3]
={\rm det}(-\ds+\rho_n)
\eeq
from which one can deduce that
\beq
{\rm det}(\ds+\rho_n)=\sqrt{{\rm det}(-\ds^2+\rho_n^2)}
\eeq
The eigenvalue equation of $-\ds^2+\rho_n^2$ is
\beq
(-D_iD_i-eB\sigma_3+\rho_n^2)\psi=\nu\psi
\eeq
Again from the Landau levels, we know that
\beq
\nu=2eB(l+1/2-s_z)+\rho_n^2
\eeq
Therefore,
\begin{eqnarray}
{\rm det}(\ds+\rho_n)&=&\frac{|eB|}{2\pi}
\sqrt{\prod_{l=0}^{\infty}(2eBl+\rho_n^2)[
2eB(l+1)+\rho_n^2]}\\
&=&\frac{|eB|}{2\pi}\rho_n\prod_{l=0}^{\infty}[2(l+1)eB+\rho_n^2]
\end{eqnarray}
which agrees with (51).\\
\indent The total effective action is then
\beq
\Gamma=\Gamma_{odd}-\frac{|eB|}{2\pi}
\sum_{n=-\infty}^{+\infty}(\ln\rho_n+\sum_{l=0}
^{\infty}\ln[\rho_n^2+2(l+1)eB])
\eeq
which is divergent. With this effective action, one can
discuss the induced particle density and the spin of the
system. But the expressions are not as simple as in the
non-relativistic case. \\
\section{Discussions}
\indent To conclude this paper, we make some
discussions.
For the background (7), the effective action can also be computed
as the zero temperature case in\cite{s19}. We here first seperate
$\Gamma$ into a parity-odd part and an even part. Both calculations should
be in accordance with each other. We know that in general
at zero temperature, the functional determinant can be
expanded in terms of the powers of $\frac{1}{m}$\cite{s29}
\beq
-i\ln{\rm det}(i\rD\pm m)=\pm W_{CS}+\frac{1}{24\pi m}\int d^3x
F_{\mu\nu}F^{\mu\nu}+O(\frac{\p^2}{m^2})
\eeq
Unfortunately, we can not make a direct comparison between (58) and (59)
because of the sum $\sum_l$. Eq(58) can be written as
\beq
\Gamma=\Gamma_{odd}-\frac{|eB|}{2\pi}\cdot 3\sum_{n=-\infty}^{\infty}
\ln\rho_n-\frac{|eB|}{2\pi}\sum_{l=0}^{\infty}\sum_{n=-\infty}^{\infty}
\ln(1+\frac{2(l+1)eB}{\rho_n^2})
\eeq
So the second term in (59) should correspond to the first term
of the expansion of $\ln(1+x)$ of the third term in (60). In the
case $\mu=\tilde{A}_3=0$, the sum over $n$ can be accomplished
using the formula
\beq
\sum_{n=-\infty}^{\infty}\frac{1}{(2n+1)^2+\theta^2}=\frac{1}{\theta}
(\frac{1}{2}-\frac{1}{e^{\theta}+1})
\eeq
But the sum over $l$ is troublesome.\\
\indent Finally, we would like to mention that apart
 from the interests explained in the Introduction, there
is another interest  relevent to bosonization. If the fermion determinants
can be calculated exactly, we may employ the duality-transformation
approach\cite{s30}
to bosonize the fermion models as in\cite{s31}.

\vskip 1in
\underline{\bf Acknowledgement}  
 This work was supported by the Funds for
Young Teachers of Shanghai Education Commitee and in part by the National
Science Foundation of China under Grant No. 19805004 .\\

\end{document}